\documentclass[letterpaper,12pt,floatsatend]{article} 
\usepackage{osajnl2} 
\usepackage{amsmath}
\usepackage{graphics}
\begin{document}

\bibliographystyle{epj}
\title{On the statistical interpretation of optical rogue waves}

\author{Miro Erkintalo and Go\"{e}ry Genty}
\address{Tampere University of Technology, Institute of Physics, Optics Laboratory, FIN-33101 Tampere, Finland}

\author{John M. Dudley}
\address{Institut FEMTO-ST, D\'{e}partment d{'}Optique P. M. Duffieux, CNRS UMR 6174, \\ Universit\'{e} de Franche-Comt\'{e}, 25030 Besan\c{c}on, France}

\address{Corresponding author: miro.erkintalo@tut.fi}

\begin{abstract}
Numerical simulations are used to discuss various aspects of ``optical rogue wave" statistics observed in noise-driven fiber supercontinuum generation associated with highly incoherent spectra.  In particular, we consider how long wavelength spectral filtering influences the characteristics of the statistical distribution of peak power, and we contrast the statistics of the spectrally filtered SC with the statistics of both the peak power of the most red-shifted soliton in the SC and the maximum peak power across the full temporal field with no spectral selection.  For the latter case, we show that the unfiltered statistical distribution can still exhibit a long-tail, but the extreme-events in this case correspond to collisions between solitons of different frequencies.  These results confirm the importance of collision dynamics in supercontinuum generation.  We also show that the collision-induced events satisfy an extended hydrodynamic definition of ``rogue wave'' characteristics.
\end{abstract}

\maketitle

\newpage
\section{Introduction}
\label{intro}

Nonlinear dynamical systems are often associated with extreme sensitivity to initial conditions and the exponential divergence of two arbitrary close inputs \cite{Scott-2007}. In the optical context a significant example of such a system is the propagation of energetic optical pulses in a highly nonlinear fiber that leads to the generation of  a white-light supercontinuum (SC) \cite{Dudley-2006}.  Indeed, it has been known for some time that noise can cause significant SC fluctuations \cite{Islam-1989a,Islam-1989b,Nakazawa-1998,Nakazawa-1999} and indeed noise-induced bifurcation-like divergence of different phase trajectories in SC generation has been demonstrated explicitly in stochastic simulations (see Fig. 18 of Ref. 2).  In most studies, however, the fundamental aspects of these noise characteristics have not been the main area of study; work has focussed more on identifying useful parameter regimes where the noise can be reduced to acceptable levels for applications such as the important area of optical frequency metrology \cite{Corwin-2003a}.

Recent studies, however, have provided significantly new insight into the characteristics of SC noise, and reinvigorated fundamental research into its origin.  These studies were initiated by an important paper by Solli\textit{ et al.} who pointed out that it was possible to directly determine the statistics of peak power fluctuations across specific wavelength bands of the broadband SC spectrum measured at the fiber output \cite{Solli-2007}.  In particular, by isolating shot-to-shot fluctuations on the long wavelength edge of the spectrum, they measured long-tailed distributions of peak power, associated with the generation of a small number of  high power events appearing as rare outcomes from an ensemble of initial conditions that were identical apart from low amplitude noise.  An analogy with the ``rogue waves'' of oceanic infamy was drawn, and these results stimulated further studies in an optical context, with numerical studies and experiments studying SC statistics over a wider parameter range, including femtosecond and continuous wave excitation \cite{Dudley-2008,Genty-2009,Erkintalo-2009,Mussot-2009}.  Other work has made links between the SC instabilities and the wider field of soliton turbulence \cite{Barviau-2008,Taki-2010,Genty-2010}.  Of course, many additional recent advances are covered in other papers in this Special Issue.

When filtering on the long wavelength edge of the output SC and measuring subsequent statistics, an association has been made between the high power events in the statistical distribution and solitons that have undergone ``larger than average'' frequency shifts  through the Raman soliton self-frequency shift (SSFS).  Indeed, some of us have previously carried out explicit studies examining the propagation dynamics of these ``rogue soliton'' events in more detail in order to attempt to determine particular characteristics underlying their extended frequency shifts \cite{Dudley-2008}. More recent work has looked further at the initial stages of soliton emergence from a phase of modulation instability where the pulse localisation can be described in terms of Akhmediev ``breather'' characteristics \cite{Akhmediev-1986,Dudley-2009}.  Interestingly, the intrinsic statistics of Akhmediev breather evolution \textit{before} the onset of any SC-like broadband spectral generation has also been the subject of study in the context of optical and hydrodynamic rogue waves \cite{Akhmediev-2009a} and the role of collisions has shown to be very important in generating statistical distributions with extended tails  \cite{Akhmediev-2009b}. Significantly, oceanic studies have also pointed out the importance of wave-crossing and interactions in generating large amplitude waves via nonlinear mechanisms \cite{Onorato-2006,Shukla-2006}.

In this paper, our goal is to further consider these aspects of the statistical analysis of SC noise for the case of high soliton number ($N > 100$) excitation in the picosecond regime.  Although some of the conclusions drawn may well be very general and apply to other regimes of noise-driven SC generation, the dynamics are complex and more case-by-case studies will likely necessary in order to develop a fully unified picture.  For our parameters,
our first main result is that whilst the tails of distributions obtained with long pass filtering can indeed include some distinct solitons that have undergone larger than average Raman frequency shifts, the filtering procedure introduces significant spectral clipping on solitons further from the SC edge.  An analysis of the peak power distribution of the most red-shifted soliton without spectral filtering remains skewed, but significantly less so than the L-shaped distribution obtained with filtering.  A related result that we obtain is that the distribution of SSFS wavelengths is Gaussian and does not exhibit long-tailed characteristics.  These results may have significant impact on the precise way the spectral filtering and associated measurements on SC spectra should be interpreted in the context of understanding soliton dynamics and studies of optical rogue waves. Our second main result concerns the statistics of the peak powers of all solitons present on the temporal profile at the fiber output without any wavelength selection at all.  Analysing the peak power distribution in this case, we consider how a commonly used definition of ``rogue waves'' used in hydrodynamics selects out a very small number of highest-amplitude pulses arising from soliton collisions occurring at the fiber output.   This study confirms earlier suggestions of the importance of collision events in SC generation \cite{Frosz-2006,Luan-2006} and in rogue soliton propagation \cite{Genty-2009}.  Indeed, it may well be that our results point to collision dynamics as a primary and general mechanism for rogue wave formation in many other systems.  Indeed, within an optical context, investigating the details of collision dynamics in SC generation is an area of current activity \cite{Mussot-2009,Erkintalo-2010}.

\section{Illustrative Numerical Results}
\label{sec:1}
Our numerical simulations are based on a generalized NLSE model which has been shown to correctly reproduce the nonlinear propagation of broadband pulses \cite{Dudley-2006}:
\begin{equation}
\frac{\partial A}{\partial z} - \sum_{k\geq2}\frac{i^{k+1}}{k!}\beta_k\frac{\partial^kA}{\partial T^k} = i\gamma\left(1+i\tau_{shock}\frac{\partial}{\partial T}\right)\bigg(A\int_{-\infty}^{\infty}R(T')|A(z,T-T')|^2dT'\bigg) \label{gnlse}.
\end{equation}
Here $A(z,T)$ represents the field envelope, and the $\beta_k$'s and $\gamma$ are the usual dispersion and nonlinear coefficients, respectively. The nonlinear response function $R(t) = (1-f_R)\delta(t)+f_Rh_R(t)$ includes both instantaneous electronic and delayed Raman contributions, with $f_R=0.18$ representing the Raman contribution. For $h_R(t)$ the experimentally measured fused silica Raman cross-section is used \cite{Dudley-2006}. The time derivative term on the right-hand side models the dispersion of the nonlinearity and is associated with such effects as self-steepening and optical shock formation, characterized by a time scale $\tau_{shock}$. In addition to spontaneous Raman noise, input shot noise is accounted for semiclassically through the addition of a noise seed of one photon per mode with random phase in each spectral discretization bin.

We first illustrate the numerical results obtained based on the injection of 150 W peak power and 5 ps duration gaussian pulses with different random noise seeds in a 25 m long photonic crystal fiber with a zero-dispersion wavelength at 1030 nm.   Fiber parameters are those used in previous studies \cite{Dudley-2008,Lafargue-2009}. Specifically, we generate an ensemble of 1000 SC events for a pump wavelength at 1035 nm slightly in the anomalous dispersion regime so that SC generation is seeded from noise-driven modulation instability.
Figure 1(a) shows the superposition of all the individual spectra from the simulated ensemble (gray traces) and the calculated mean spectrum (solid line) at the fiber output. When we use long-pass filter positioned on the long wavelength edge of the SC spectrum (1260~nm for the results shown), we can extract a time series from the Fourier Transforms of the filtered spectral components, and perform statistical analysis of the power of the filtered temporal peaks.  Figure 1(b) shows the intensity distribution histogram that results,  and we can see that it possesses an L-shape with a long tail characteristic of extreme-value events and similar to the skewed statistical distributions of wave heights observed in oceanic rogue waves \cite{Pelinovsky-2008}.

\begin{figure}[h]
\centering
\resizebox{.7\columnwidth}{!}{%
\includegraphics{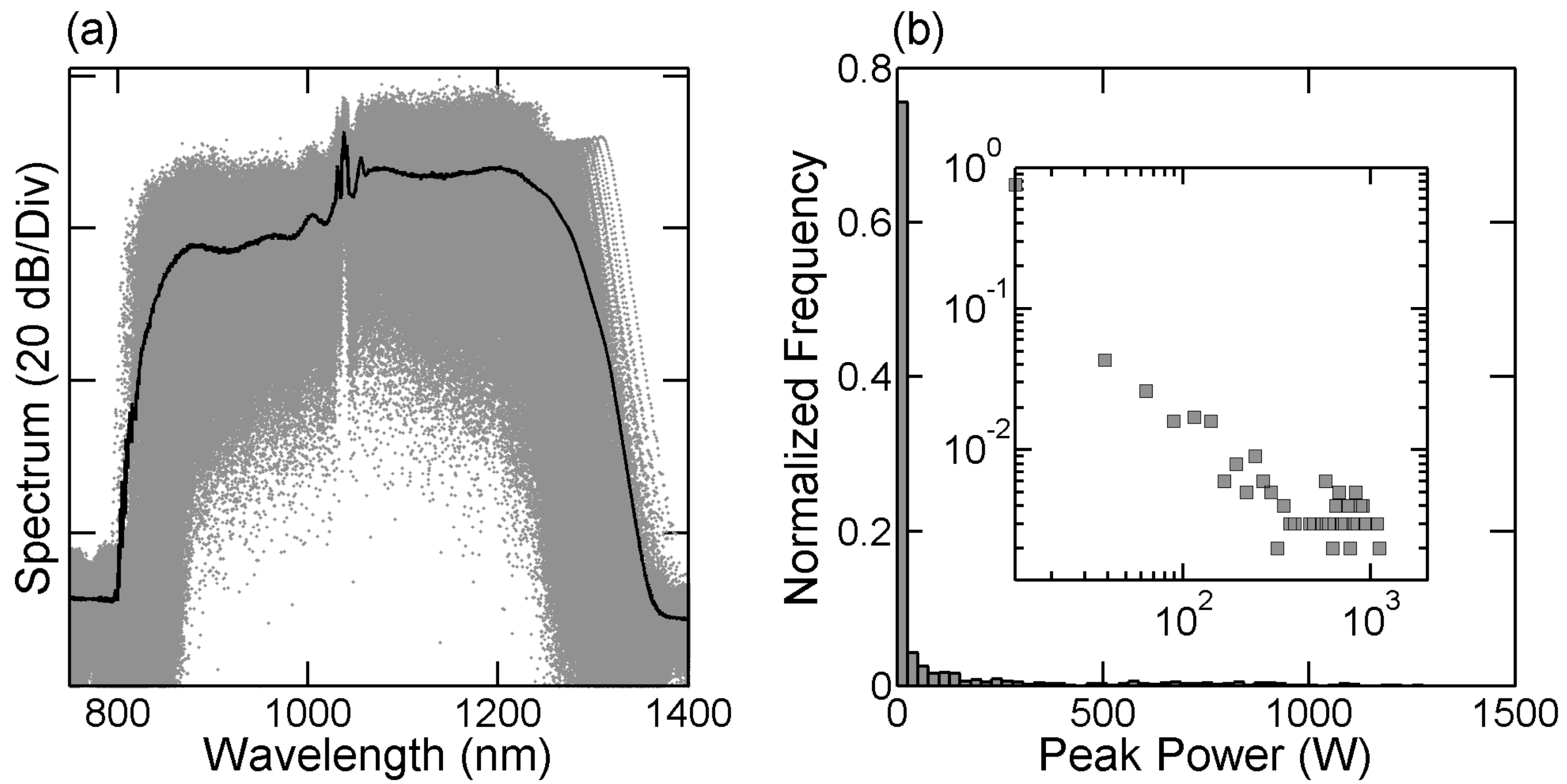}}
\caption{(a) Output spectra of 1000 individual simulations (gray) and the calculated mean spectrum (black). (b) Histogram of the intensity distribution when each simulation spectrum is filtered with a long-pass filter at 1260 nm. The inset uses a log-log representation.}
\label{fig:1}
\end{figure}

\section{Statistics with Spectral Filtering}
\label{sec:2}

 The use of a spectral filter will fully capture all the solitons that undergo significant Raman SSFSs well-beyond its long pass wavelength such that the effect of the filter on the soliton spectrum is negligible.  On the other hand, for any given realisation in the ensemble, the most red-shifted soliton in the output SC supercontinuum may not in general lie completely beyond the filter transmission wavelength, and thus will be significantly clipped by the filter.  In this case, peak power of the corresponding Fourier-transformed peak in the filtered time series will not accurately reflect the peak power of the original unfiltered red-shifted soliton.

 We show this quantitatively in Fig. 2 where we have performed a detailed inspection of the 1000 events in the ensemble to determine the exact peak power of the most red-shifted soliton in each realisation without any spectral filtering.    This was accomplished by using a two-dimensional filtering procedure based on the calculated spectrogram of the field to isolate the longest wavelength soliton cleanly and then to determine its peak power.  This approach also removes the influence of normal dispersion dispersive wave radiation components from the corresponding temporal soliton profile allowing accurate determination of the peak power.  The results of this analysis are shown in Fig. 2~(a) where we plot the histogram of the most red-shifted soliton without filtering.  In this case, the distribution remains skewed but does not possess the very strong L-shape as in Fig. 1~(b).  Interestingly, however, the skewed form of the distribution is still fitted by a Weibull distribution that is commonly used to fit extreme value processes.  On the other hand, other distributions such as the Rayleigh distribution can also provide good fits to the statistics for this case.  We have also performed an analysis of the wavelength distribution of the most red-shifted solitons in the SC spectra and these results are shown in Fig. 2(b).  Here we plot the distribution of soliton frequency shifts (relative to  the pump) and we find that the distribution is in fact near-Gaussian, although there is a very small residual skewness.

 An important conclusion to draw from these results is that, with the parameter regime we consider here, the SC statistics obtained with spectral filtering do not provide a correct quantitative measure of the properties of the most redshifted soliton when examining an ensemble of noisy SC spectra.  Our results also show that it is not as a result of extreme value statistics in the Raman SSFS that skewed distributions are measured in the corresponding (filtered or unfiltered) soliton peak powers.  On the other hand, spectral filtering is a very convenient experimental technique and will certainly capture a certain number of discrete high power soliton events.  Also, in experiments where filtering is used to select a portion of the spectra for particular applications (eg: spectroscopy) then the statistics measured after filtering (whatever distribution they may take) are indeed those that are relevant.  But for more general considerations involving soliton dynamics, it is important to carefully consider the effect of spectral filtering in clipping the spectral content of many of the solitons that one wishes to characterise.

\begin{figure}[h]
\centering
\resizebox{.7\columnwidth}{!}{%
\includegraphics{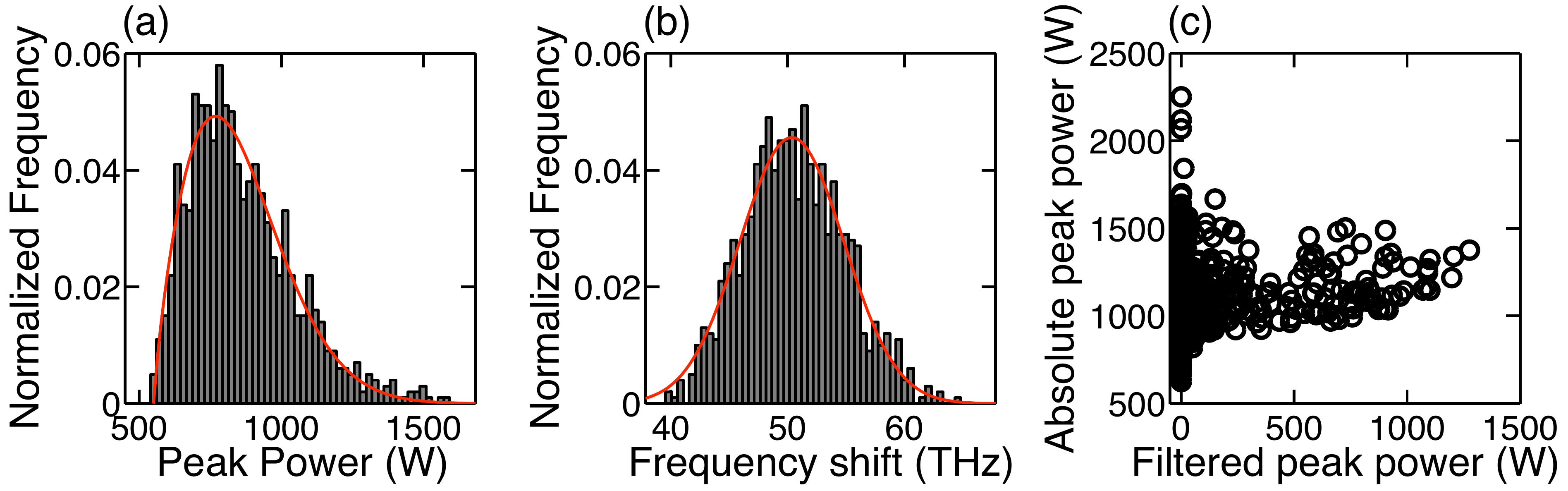} }
\caption{(a) Statistical distribution of the intensity of the most redshifted soliton at the fiber output in the absence of spectral filtering together fitted with a Weibull distribution (in red). (b) Statistical distribution of the (absolute value of) frequency-shift relative to the pump taken at the -20 dB bamdwidth of the most redshifted soliton at the fiber output in the absence of spectral filtering fitted with a Gaussian distribution (in red). (c) Plot of soliton peak power without filtering (``Absolute Peak Power'') against the peak power measured when filtering at cutoff wavelengths of 1260 nm.}
\label{fig:2}       
\end{figure}

\section{Statistics with Full Field Analysis}
\label{sec:3}

Of course, when considering the dynamics in general terms, there is no \textit{a priori} reason to restrict the analysis only to the most red-shifted solitons in the generated SC spectra.  It is equally meaningful to consider the maximum peak power observed across a given output temporal profile of the SC without any filtering or processing at all and indeed, this may well be a more appropriate approach when considering analogies with the oceanic environment.  So with this method, we analyse the temporal profile of the SC field including all contributing frequency components in the spectrum (except for some low amplitude normal dispersion regime components), and we search for the "absolute maximum peak power" for each realization in the ensemble.  We refer to this as ``Full Field Analysis''.

Firstly, it is interesting to compare how the absolute maximum peak power from the full temporal intensity profile correlates with the peak power measured in the presence of a spectral filter.  These results are shown in Figs. 2(c) for a filter cutoff wavelength at 1260~nm corresponding to a filter close to the long wavelength edge.  We see that although there is a loose correlation above filtered peak powers of 500~W, there is a very high degree of scatter and indeed, the full temporal profile contains some events with absolute peak power exceeding 2 kW which are not captured at all by the spectral filtering.  In fact, these events also do not appear in the histogram in Fig. 2(a) calculated using the spectrogram approach to isolate the extreme red-shifted solitons, suggesting that they have a different physical origin that we now discuss in detail.

In order to obtain quantitative information on the relative amplitudes of all the high-amplitude peaks present on the temporal profile we employ a numerical peak-detection algorithm.  We consider only high power pulses in the anomalous dispersion regime where strongly localised pulses and solitons are expected, so our approach includes filtering out spectral content in the normal dispersion regime. We subsequently detect the local maxima in the temporal domain and then detect the neighboring zeros of the maximum (or values close to zero - we used a threshold of $<$ 5\%), which allows to isolate individual high power pulses.  In fact, analysis of the pulse characteristics (eg: soliton number) isolated with this technique confirms that they are indeed solitons. Note that this method can be envisaged as an adaptation of the zero-crossing technique commonly used in hydrodynamics to isolate individual wave events.  We neglected all waves/pulses with a peak power lower than 100 W as these typically correspond to non-solitonic radiation and residual pump components typically consisting of weakly modulated wave train and not a localised pulse.  Although the choice of this threshold may seem arbitrary it is not possible to determine an absolute threshold value that would allow to reject all -and only- non-solitonic radiation content. Nonetheless, we checked that with this choice of threshold only very few solitons with low peak power were omitted and therefore the statistics of the nearly 10000 captured solitons are not significantly affected.

Figure 3(a) illustrates the statistical distribution of all the soliton amplitudes obtained from nearly 10000 individual events at the fiber output.  This corresponds to an average of 10 solitons per shot emerging from the breakup of the input pulse. The distribution is seen to exhibit a rather complex shape with two distinct maxima and a long tail indicating that very few solitons exhibit a significantly larger amplitude than the mean.

To progress further, we now introduce a hydrodynamic definition of the events that constitute ``rogue behaviour''.  Specifically, we define an optical rogue event here as a soliton whose height (peak power) is more than twice the significant wave height (i.e. the mean peak power of the one-third largest amplitude solitons). Remarkably, with this choice of definition and as emphasized in Fig. 3(b), only \textit{three waves} located at the extreme end of the distribution tail satisfy this criterion for consideration as an optical rogue wave.  Even perhaps more surprisingly: (i) the shots to which these waves correspond lie in the zero-intensity bin in Fig. 1(b) when long-wavelength spectral filtering is used and (ii) none of these correspond to the most redshifted soliton of the associated shot and are thus absent in Fig. 2(a). Significantly, this means that considerations of optical rogue waves in terms of events that survive after spectral filtering or that correspond to extreme redshifting trajectories actually exclude these particular large amplitude events that correspond closely to a hydrodynamic definition.

\begin{figure}[t]
\centering
\resizebox{.7\columnwidth}{!}{%
\includegraphics{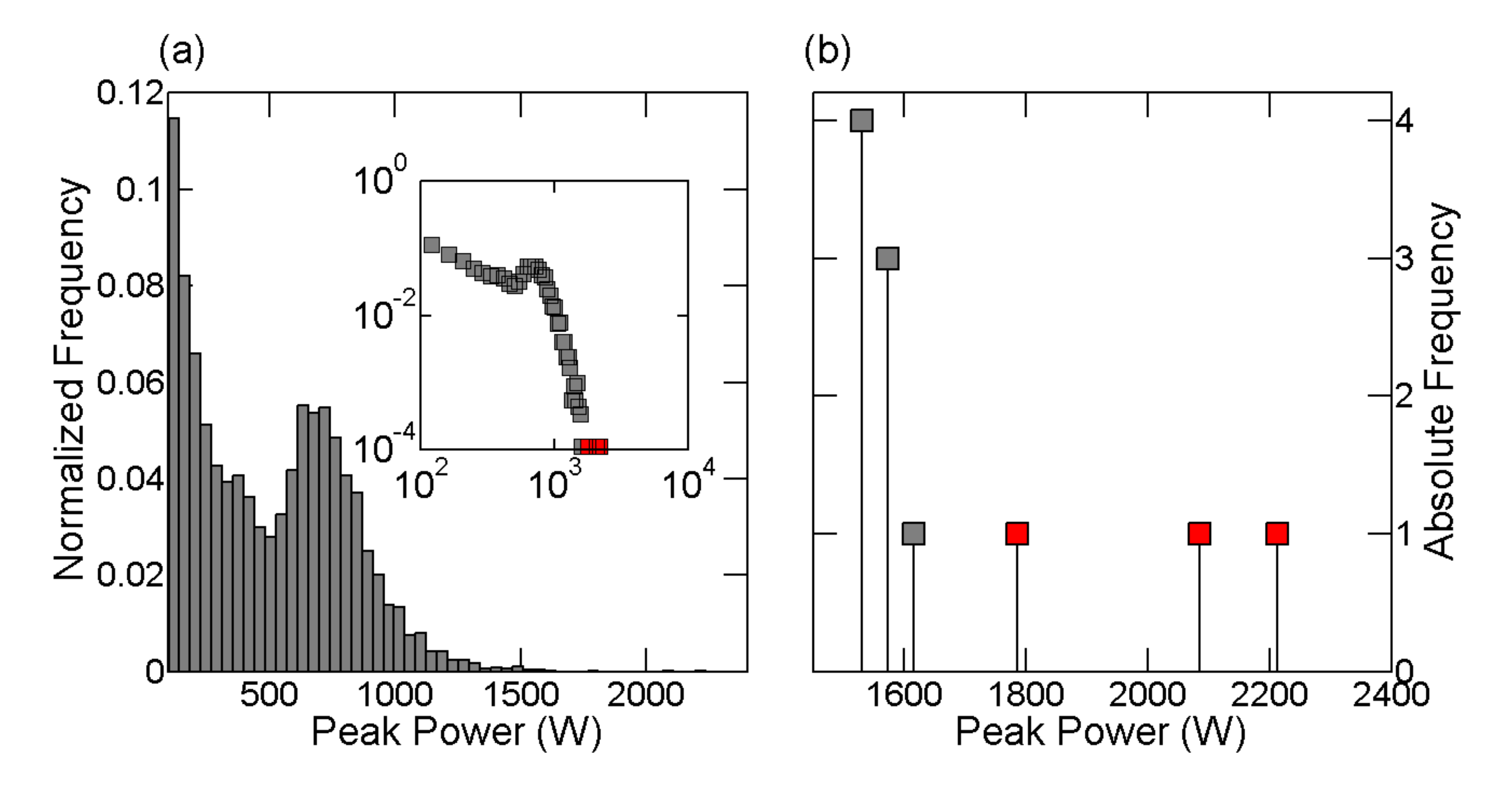} }
\caption{(a) Statistical distribution of all soliton amplitudes at the fiber output. The inset plots the results on a log-log scale. (b) Zoom on the tail emphasizing the events that qualify as ``rogue'' using hydrodynamic definitions (red).}
\label{fig:3}       
\end{figure}

To see why these highest amplitude events are missed by these two approaches to detection, we compare the time-frequency spectrograms of two particular events in Fig. 4.  Specifically, Fig. 4(a) plots an event where a single soliton experiences significant red-shift beyond the cutoff-wavelength of the long-pass filter (the dashed line in the figure). Its detection therefore yields a peak power of around 1.3 kW.  On the other hand, Fig. 4(b) shows that the realisation with the maximum absolute peak power (exceeding 2 kW in the time trace projected below the spectrogram) arises physically from the nonlinear superposition or collision between two solitons with different wavelengths.  Significantly, because this case does not yield significant spectral content beyond the cutoff wavelength, spectral filtering would only detect it as a low power event.   In addition, even consideration of the single most red-shifted soliton peak power from the spectrum would also not accurately represent the physical nature of this particular event.  It is necessary to consider such collision-events as distinct from single soliton events in order to understand their contribution to the rogue wave population. In this regard, we stress that we have checked that similar collision scenarios occur for the two other events in Fig. 3 that satisfy our hydrodynamic-based definition of optical rogue waves.

\begin{figure}[p]
\centering
\resizebox{0.6\columnwidth}{!}{%
\includegraphics{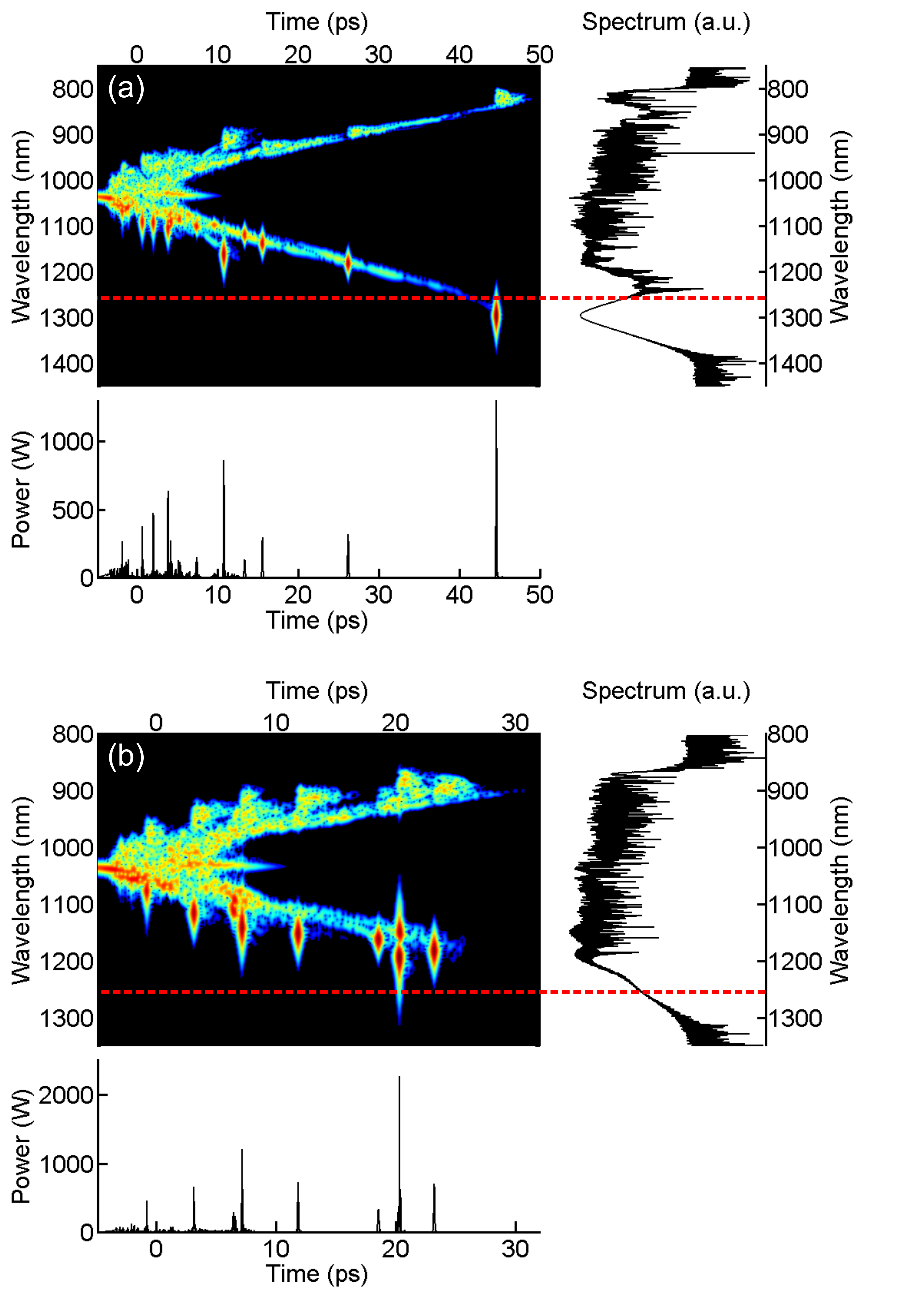} }
\caption{(a)Spectrogram representation of the output field for the event associated with the maximum peak power for filtered statistics and (b) overall maximum peak power of the full field. The spectrograms were calculated from the output fields assuming a 200 fs gate function. Red vertical line indicates the cutoff wavelength of the long-pass filter (1260 nm).}
\label{fig:4}       
\end{figure}

\section{Discussion: the Subtle role of Collisions}
\label{sec:4}

As mentioned above, restricting the analysis of temporal content in the SC field to red-shifted solitons or to the filtered long wavelength spectral edge can actually miss the highest amplitude collision-induced peaks on the output temporal profile.  And significantly, it is these peaks which actually satisfy an equivalent hydrodynamic definition as a ``rogue'' event.

The relationship between collisions and optical rogue waves, however, is rather subtle, because collisions are also very important in understanding the class of extreme red-shifted rogue solitons captured with spectral filtering at the SC long wavelength edge.  To discss this in detail, Fig. 5 shows the dynamical evolution of the pulse envelope in both spectral and temporal domains for a typical event in the tail of the intensity distribution shown in Fig. 1(b).  In this case, it can be seen how after 14~m of propagation a localized structure is suddenly ejected from the long wavelength edge of the SC spectrum. In the time domain this corresponds to a collision between solitons following different trajectories because of their different group velocities.

The effect of the collision at 14~m is two-fold (i) it transfers energy to the soliton with lower frequency, which in turn induces a subsequent change in its velocity (i.e. increases the rate of Raman shifting) and (ii) it induces a cross-frequency shift onto the largest amplitude soliton towards shorter frequencies \cite{Kumar-1993}. More specifically, during the collision, the low frequency soliton overlaps with the Raman gain induced by the high frequency soliton leading to significant exchange of energy. As a result, the peak power of the low frequency soliton is increased significantly leading to a corresponding enhancement in the Raman self-frequency shift after the collision point.  In addition, if the frequency difference between the colliding solitons is less than 13.2 THz so that the low frequency soliton coincides with the positive slope of the Raman gain induced by the high frequency soliton, the energy exchange leads to a net shift of the low frequency soltion during the collision process.

This mechanism is knows as soliton cross-frequency shift which, in contrast to the self-frequency shift,  is a localized effect occurring only over the distance the colliding solitons temporally overlap \cite{Agrawal-2006}. The relative magnitude of the cross-frequency shift depends on several factors such as the amplitudes of the colliding solitons and their frequency separation. In general, the larger (smaller) the amplitudes (frequency separation) of the colliding solitons, the larger the cross frequency shift \cite{Malomed-1991,Lakoba-1999}. As a result of the cross-frequency shift and energy exchange during the collision the rate of subsequent self-frequency shift for the low frequency soliton is effectively enhanced compared to the case where no collision would occur. Therefore, it is primarily the collision process that leads eventually to an extension of the SC bandwidth at the fiber output.

\begin{figure}[t]
\centering
\resizebox{.7\columnwidth}{!}{%
\includegraphics{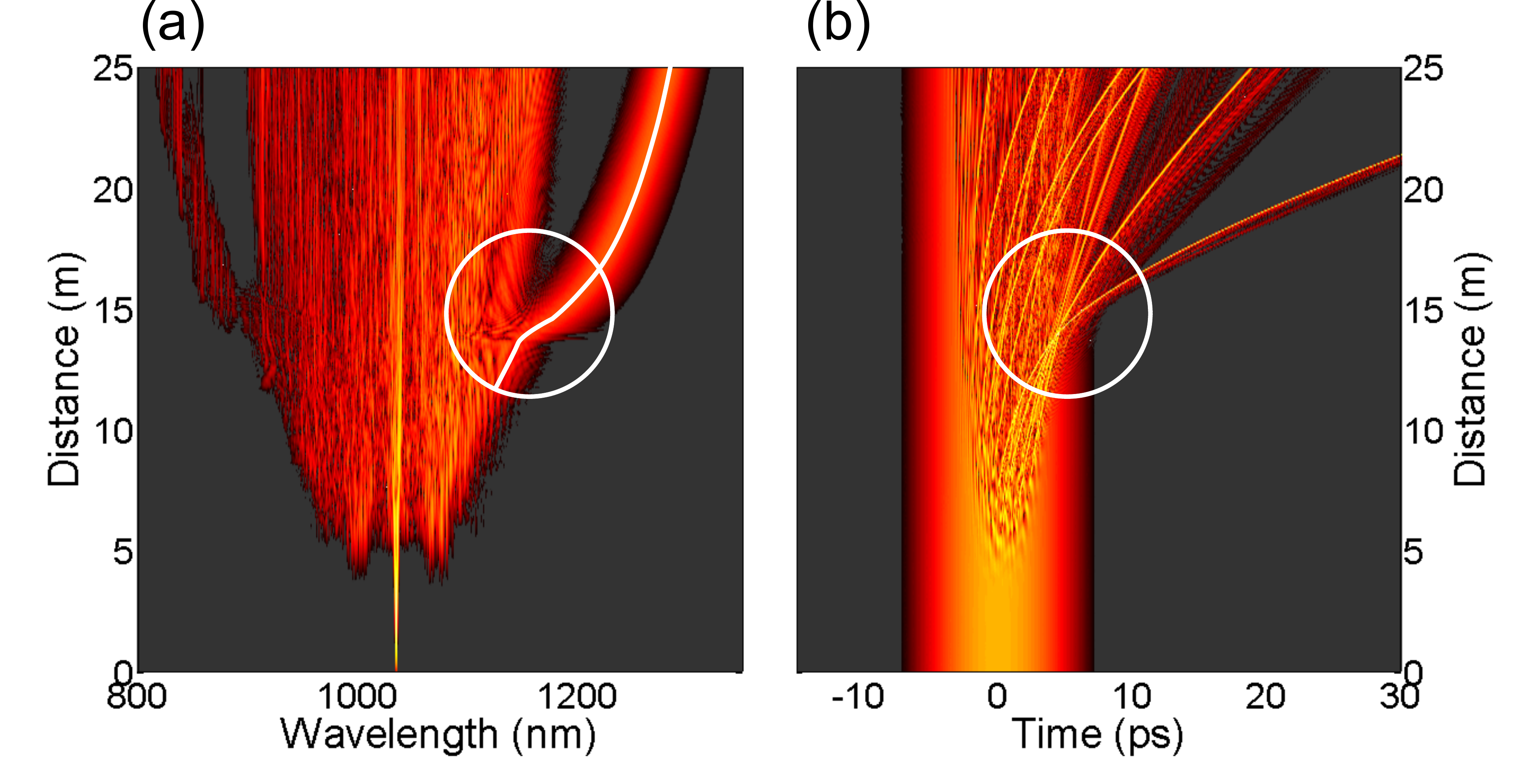} }
\caption{Density plots of (a) spectral and (b) temporal evolution for a traditional optical rogue wave. Point of collision and the following transient behaviour in the spectral evolution is emphasized with white curves.}
\label{fig:5}       
\end{figure}

It can now be appreciated that whether or not a soliton is fully captured by a spectral filter at the fiber output is mostly determined by the occurrence of a collision that induces a large cross-frequency shift and significant energy exchange. This is in fact confirmed when carefully inspecting all the events corresponding to the tail of the distribution shown in Fig. 1(b) and whose evolution is found to follow similar dynamics. Furthermore, solitons that are fully captured by the spectral filtering have experienced a significant frequency-shift and, due to the positive third-order dispersion coefficient, their amplitude has significantly dropped after the collision. On the other hand the solitons that are partly or not captured by the spectral filter have experienced a reduced red-shift and their amplitude may be larger than that of the fully-captured solitons.   The important conclusion here is that the histogram of the filtered intensities as shown in Fig. 1(b) should not be interpreted as describing the actual power distribution of the most red-shifted solitons but rather the distribution of SC spectra exceeding the filter cutoff wavelength via soliton collision-induced mechanisms as described above. In this regard, note that although collision-related dynamics lead to extended SC spectrum, the positive third-order dispersion tends to reduce the rate of red-shift so that the statistical distribution of SC spectra is in fact nearly Gaussian (see Fig. 2(b)).

We now consider the properties of high power events analysed along the propagation distance.  In particular, because collisions are localized events, it is very likely that the most powerful events occuring in each shot - and along the full propagation distance - differs from that captured at a given distance. This is precisely what is seen in Fig. 6 which plots the distance at which the maximum amplitude reached in each realization occurs. Although no obvious systematic correlation can be extracted from the figure, of significant importance is the fact that the most powerful events mostly occurs after a distance of 10-14 m whereas it is only in rare cases that the highest peak power events generally occur at the fiber output; recall that the case above recorded only 3 collisions out of 10000 at the output.

But even in the cases where the collisions occur at the output, the maximum amplitude is still substantially reduced compared to cases where the most powerful event occurs at an early stage. Both these observations can in fact be explained by considering the dynamics leading to the development of the SC. After the initial phase of MI and soliton emergence, the solitons subsequently begin to experience collisons due to different group velocities (resulting from the higher-order dispersion and Raman scattering perturbations.) It is precisely at this stage where the solitons start colliding that corresponds to the 10-14 m distance range (of course this depends on the particular parameters chosen). And since the solitons have not propagated over a sufficient distance so as to be well-separated from each other in time it is therefore not surprising that the probability of observing large amplitude events is substantially larger than after a longer propagation distance. Indeed, after longer distances solitons are well separated in time and posses a lower amplitude due to the effect of positive third-order dispersion, which significantly decreases the probability of observing large amplitude collisions. This explains why only in rare cases the most powerful events in each shot are in fact not captured for distances where the SC has fully developed.

\begin{figure}[t]
\centering
\resizebox{.7\columnwidth}{!}{%
\includegraphics{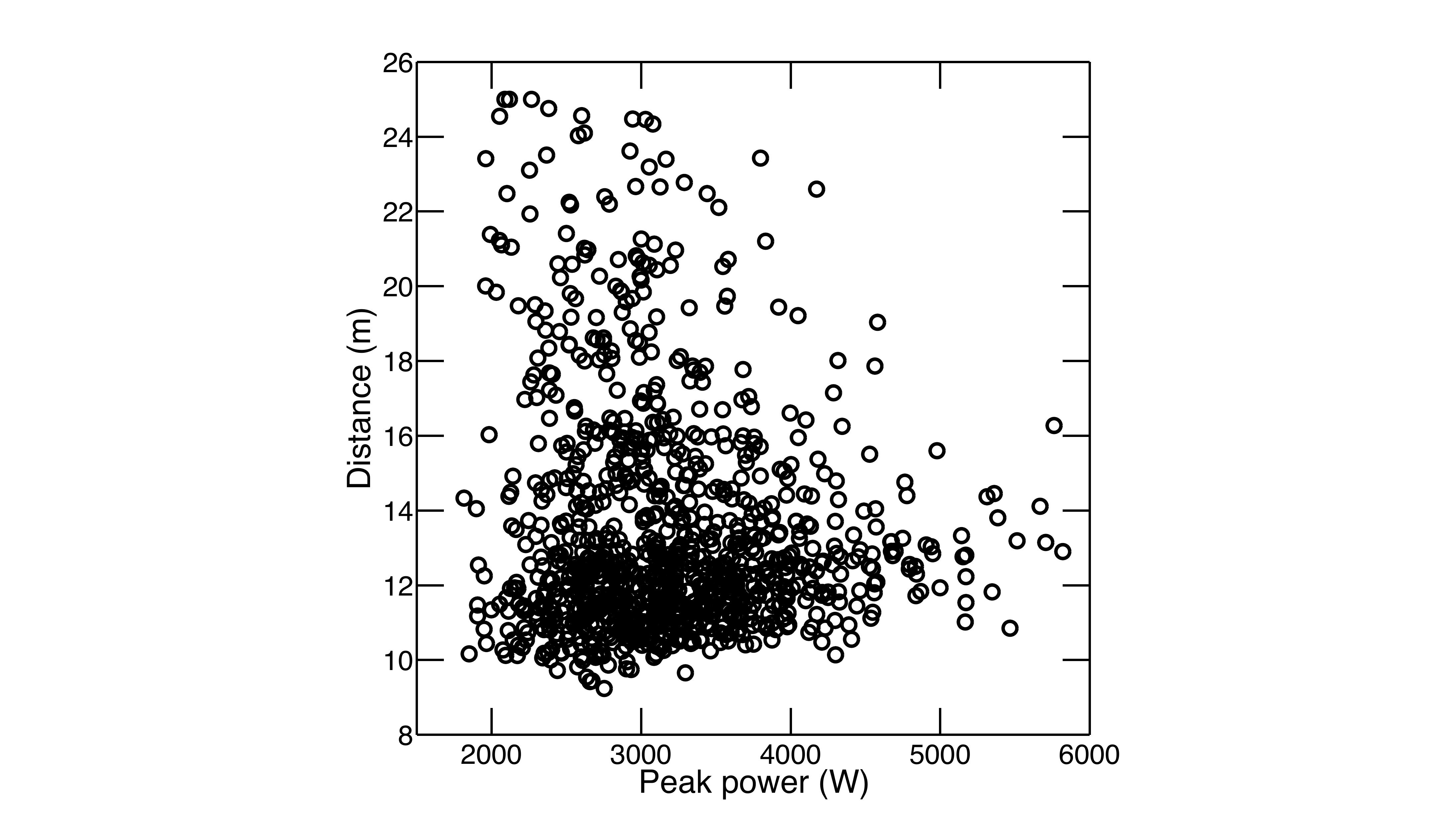} }
\caption{Propagation distance at which the maximum intensity attained during each shot is reached.}
\label{fig:6}       
\end{figure}

\section{Conclusions}
\label{sec:5}

In this paper, we have considered several different aspects of how the statistics of high intensity and/or long wavelength ``events'' in SC generation should be analysed for the regime of high soliton number noise-driven picosecond excitation.  Firstly, the use of a spectral filter will fully capture all the solitons  at frequencies beyond its long pass wavelength but will clip (or not measure at all) solitons at wavelengths below the filter transmission wavelength. In this case, the power of peaks in the filtered time series will not accurately reflect the peak power of the original unfiltered red-shifted solitons generated in the SC.  Moreover, spectral clipping in the vicinity of the filter transmission wavelength can lead to some soliton pulses being attenuated to the point when they contribute significantly to the portion of the L-shaped distribution containing low power events.   The events in the tails still do correspond to distinct soliton pulses that have undergone larger than average frequency shifts due to collision-related mechanisms (eg: the Raman cross-frequency shift), but the association of soliton dynamics with an extreme L-shaped distribution must be considered with care because the vertical low power branch of the "L" can actually contain a significant population of clipped soliton events.

A potentially clearer association between SC fluctuations and hydrodynamic rogue wave statistics appears when once considers a full field analysis.  This reveals the presence of a very small number of collision-induced events occurring at the output of the fiber.  These do not correspond to extreme red-shifting trajectories, but nevertheless yield a higher peak power than the single soliton events that do shift to longer wavelengths.  They also occur more rarely and can be related more rigorously with the hydrodynamic rogue waves through an adapted hydrodynamic definition.  We have also considered in detail the physics of collision processes in determining the propagation dynamics that yield both ``filtered'' and ``full field'' classes of optical rogue wave event. Finally, we remark that the dynamics of SC generation are very complex, and the conclusions drawn here must be considered within the framework of the high soliton number excitation condition, the broadband noise source used and the near-zero dispersion wavelength pumping.  Although we have fully confirmed that our conclusions apply to high soliton number excitation of picosecond pulses and to high power continuous wave pumping, other initial conditions may yield different statistics and more case-by-case studies will be necessary to establish a more complete picture of the statistical properties of optical rogue waves.  As in the oceanic case, understanding the statistics seems to be a central and necessary area of study in order to develop the most general interpretation possible.

\section*{Acknowledgments}
We thank the Academy of Finland (research grants 130099 and 132279), the Institut Universitaire de France, and the French Agence Nationale de la Recherche project MANUREVA (ANR-08-SYSC-019) for support. M. Erkintalo acknowledges support from the graduate school of Tampere University of Technology.


\begin{thebibliography}{30}

\bibitem{Scott-2007}
A.W. Scott, \emph{The Nonlinear Universe} (Springer, Berlin, 2007)

\bibitem{Dudley-2006}
J.M. Dudley, G.~Genty, S.~Coen, Rev. Mod. Phys. \textbf{78}, 1135 (2006)

\bibitem{Islam-1989a}
M.N. Islam, G.~Sucha, I.~Bar-Joseph, M.~Wegener, J.P. Gordon, D.S. Chemla, Opt.
  Lett. \textbf{14}(7), 370 (1989)

\bibitem{Islam-1989b}
M.N. Islam, G.~Sucha, I.~Bar-Joseph, M.~Wegener, J.P. Gordon, D.S. Chemla,
  J.~Opt. Soc. Am.~B \textbf{6}(6), 1149 (1989)

\bibitem{Nakazawa-1998}
M.~Nakazawa, K.R. Tamura, H.~Kubota, E.~Yoshida, Opt. Fib. Tech. \textbf{4}(2),
  215 (1998)

\bibitem{Nakazawa-1999}
M.~Nakazawa, H.~Kubota, K.R. Tamura, Opt. Lett. \textbf{24}(5), 318 (1999)

\bibitem{Corwin-2003a}
K.L. Corwin, N.R. Newbury, J.M. Dudley, S.~Coen, S.A. Diddams, K.~Weber, R.S.
  Windeler, Phys. Rev. Lett. \textbf{90}, 113904 (2003)

\bibitem{Solli-2007}
D.R. Solli, C.~Ropers, P.~Koonath, B.~Jalali, Nature \textbf{450}, 1054 (2007)

\bibitem{Dudley-2008}
J.M. Dudley, G.~Genty, B.J. Eggleton, Opt. Express \textbf{16}, 3644 (2008)

\bibitem{Genty-2009}
G.~Genty, J.M. Dudley, B.J. Eggleton, Appl. Phys.~B \textbf{94}, 187 (2009),
  original publication: arXiv:0809.2388v1, 14 Sept. 2008

\bibitem{Erkintalo-2009}
M.~Erkintalo, G.~Genty, J.M. Dudley, Opt. Lett. \textbf{34}, 2468 (2009), ISSN
  {0146-9592}

\bibitem{Mussot-2009}
A.~Mussot, A.~Kudlinski, M.~Kolobov, E.~Louvergneaux, M.~Douay, M.~Taki, Opt.
  Express \textbf{17}, 17010 (2009)

\bibitem{Barviau-2008}
B.~Barviau, B.~Kibler, S.~Coen, A.~Picozzi, Opt. Lett. \textbf{33}, 2833 (2008)

\bibitem{Taki-2010}
M.~Taki, A.~Mussot, A.~Kudlinski, E.~Louvergneaux, M.~Kolobov, M.~Douay, Phys.
  Lett. A \textbf{374}, 691 (2010)

\bibitem{Genty-2010}
G.~Genty, C.~de~Sterke, O.~Bang, F.~Dias, N.~Akhmediev, J.~M. Dudley, Phys. Lett.
  A \textbf{374}, 989-996 (2010)

\bibitem{Akhmediev-1986}
N.~Akhmediev, V.I. Korneev, Theor. Math. Phys. \textbf{69}, 1089 (1986)

\bibitem{Dudley-2009}
J.M. Dudley, G.~Genty, F.~Dias, B.~Kibler, N.~Akhmediev, Opt. Express
  \textbf{17}, 21497 (2009)

\bibitem{Akhmediev-2009a}
N.~Akhmediev, A.~Ankiewicz, M.~Taki, Phys. Lett. A \textbf{373}, 675 (2009)

\bibitem{Akhmediev-2009b}
N.~Akhmediev, J.M. Soto-Crespo, A.~Ankiewicz, Phys. Lett. A \textbf{373}, 2137
  (2009)

\bibitem{Onorato-2006}
M.~Onorato, A.R. Osborne, M.~Serio, Phys. Rev. Lett. \textbf{96}, 014503 (2006)

\bibitem{Shukla-2006}
P.K. Shukla, I.~Kourakis, B.~Eliasson, M.~Marklund, L.~Stenflo, Phys.
  Rev. Lett., \textbf{97} 094501 (2006)

\bibitem{Frosz-2006}
M.H. Frosz, O.~Bang, A.~Bjarklev, Opt. Express \textbf{14}(20), 9391 (2006)

\bibitem{Luan-2006}
F.~Luan, D.V. Skryabin, A.V. Yulin, J.C. Knight, Opt. Express \textbf{14}(21),
  9844 (2006)

\bibitem{Erkintalo-2010}
M.~Erkintalo, G.~Genty, J.M. Dudley, Opt. Lett. \textbf{35}(5), 658 (2010)

\bibitem{Lafargue-2009}
C.~Lafargue, J.~Bolger, G.~Genty, F.~Dias, J.M. Dudley, B.J. Eggleton,
  Electron. Lett. \textbf{45}, 217 (2009)

\bibitem{Pelinovsky-2008}
E.~Pelinovsky, C.~Kharif, eds., \emph{Extreme Ocean Waves} (Springer, Berlin,
  2008)

\bibitem{Kumar-1993}
S.~Kumar, A.~Selvarajan, G.V. Anand, Opt. Commun. \textbf{102}, 329 (1993)

\bibitem{Agrawal-2006}
G.P. Agrawal, \emph{Nonlinear Fiber Optics}, 4th~edn. (Academic Press, San
  Diego, 2006)

\bibitem{Malomed-1991}
B.A. Malomed, Phys. Rev.~A \textbf{44}, 1412 (1991)

\bibitem{Lakoba-1999}
T.I. Lakoba, D.J. Kaup, Opt. Lett. \textbf{24}, 808 (1999)

\end{thebibliography}

 \providecommand{\micron}{\ifmmode\mu{\mathrm m}\else$\mu{\mathrm m}$\fi}
  \providecommand{\singleletter}[1]{#1} \providecommand{\hideforsort}[1]{}

\end{document}